\documentclass[sigconf]{acmart}

\AtBeginDocument{%
  }

\copyrightyear{2025}
\acmYear{2025}
\setcopyright{acmlicensed}
\acmConference[PEARC '25]{Practice and Experience in Advanced Research Computing}{July 20--24, 2025}{Columbus, OH, USA}
\acmBooktitle{Practice and Experience in Advanced Research Computing (PEARC '25), July 20--24, 2025, Columbus, OH, USA}
\acmDOI{10.1145/3708035.3736074}
\acmISBN{979-8-4007-1398-9/2025/07}

\begin{document}

\title{Parallel Ray Tracing of Black Hole Images Using the Schwarzschild Metric}

\author{Liam Naddell}
\email{liam.naddell@mail.utoronto.ca}
\author{Marcelo Ponce}
\email{m.ponce@utoronto.ca}
\affiliation{%
  \institution{Department of Computer and Mathematical Sciences, University of Toronto Scarborough}
  \city{Toronto}
  \state{ON}
  \country{Canada}
}


\begin{abstract}
Rendering images of black holes by utilizing ray tracing techniques is a common
methodology employed in many aspects of scientific and astrophysical visualizations.
Similarly, general ray tracing techniques are widely used in areas related to computer graphics.
In this work we describe the implementation of a parallel open-source program that can ray trace images in the presence of a black hole geometry.
We do this by combining a couple of different techniques usually present in parallel scientific computing,
such as, mathematical approximations, utilization of scientific libraries, shared-memory and distributed-memory parallelism.


\end{abstract}




\maketitle


\section{Introduction}
\label{sec:intro}

Ray tracing is a foundational technique in computer graphics, which can be used
to render photorealistic scenes.
A good introduction to the main techniques and practical implementations can be found
in Ref.\cite{raytracing_in_one_weekend}.
In general this implies expensive processes and requires the corresponding appropriate compute time
as well as complex and sophisticated algorithms.
The applications \cite{Peddie2019_appns} of which range from generation of high-quality realistic visualizations to
rendering of movies or animations, and up to even more exotic images of black holes,
such as the ones remarkably done for the black hole residing at the center of the galaxy M87 \cite{M87_EHT_i}
and the supermassive black hole Sagittarius A* at the center of our own galaxy \cite{SagA_EHT_i}.
In these cases, ray-tracing techniques have been applied to
General Relativistic Magneto-HydroDynamics simulations in order to produce 
the so-called fiducial templated models of the event horizon images.
Techniques such as these have allowed us to produce images and media that have captivated the minds of millions.

We should emphasize that there exist a large variety of ray tracer implementations already 
\cite{imbens2023graphicalprocessinggeodesicpropagation,10.2312/EGPGV/EGPGV12/051-060,7539599_OSPRay}
and even ones with direct astrophysical applications such as the rendering of black hole surroundings
\cite{10.2312:vmv.20221208,sharma2023mahakalapythonbasedmodularraytracing,James_2015}.
However, our approach seeks to highlight some specific features:
i) its open-source implementation;
ii) the minimalistic and simplistic implementation strategy, i.e. by tackling
mostly the actual mathematical problem using specialized libraries
to solve the governing differential equation of the problem;
iii) the ease of implementation in a fairly efficient and high-performing
way by employing standards in shared-memory and distributed-memory paradigms;
and iv) the hardware agnostic approach, i.e. not depending on specialized hardware such as GPUs.\cite{Peddie2019_hardware}

In order to model a simple black hole scenario, we can use the Schwarzschild metric \cite{schw_soln-2007}.
This models black holes which, among other properties, are non-spinning and not charged,
meaning that their deformation of the trajectories of light rays do not depend on the angle of approach.
It has been well established \cite{gravitation-mtw} that the ordinary differential equation (ODE),
\begin{equation}
	u'' - u = 3 M u^2
	\label{eq:Sch-light-ray}
\end{equation}
relates the trajectory of a light ray to the mass $M$ of a black hole, from the Schwarzschild metric;
where $u$ represents $1/r$, $r$ being the Euclidean distance of a point on the light ray to the black hole center.
$u$ is differentiated with respect to $\phi$, the azimuthal angle between the point and the black hole origin.
See Fig.~\ref{fig:nextpoint} and its accompanying section for more detail on the geometry of this equation.
After providing initial conditions,  it is possible to solve Eq.(\ref{eq:Sch-light-ray}) and trace the entire trajectory of the ray with high fidelity.

After being able to compute the trajectory of individual rays, we can use these rays to generate a full image using ray tracing. Ray tracers render a simulated scene by casting light rays from a simulated camera, and computing what objects in the scene would be visible to that ray. Hypothetically, if a particular camera would produce an image of $1920 \times 1080$ pixels, we could build the image that camera would record by casting a light ray through each pixel. Because ray tracing's unit of perception is how simulated light rays interact with a simulated scene, it is a natural choice for simulations which involve the bending or distortion of light.



\section{Implementation}
\label{sec:impl}

\subsection{Approximating the trajectory of a light ray as a piecewise function}
While the fundamental idea of taking a ray tracer and modifying it so that light rays are distorted is sound, the implementation of such an idea is fraught with difficulty. The first major challenge is that ray tracers need linear equations to be able to compute object intersection in a scene. This means we need to find a way to describe our light ray as a piecewise-defined set of linear equations. We found the best way to do this was to compute discrete points on the light ray, and compute intersection using the line segments defined by said points.

\subsection{Using GSL for ODE approximation}
\begin{figure}[h]
  \centering
  \includegraphics[width=\linewidth]{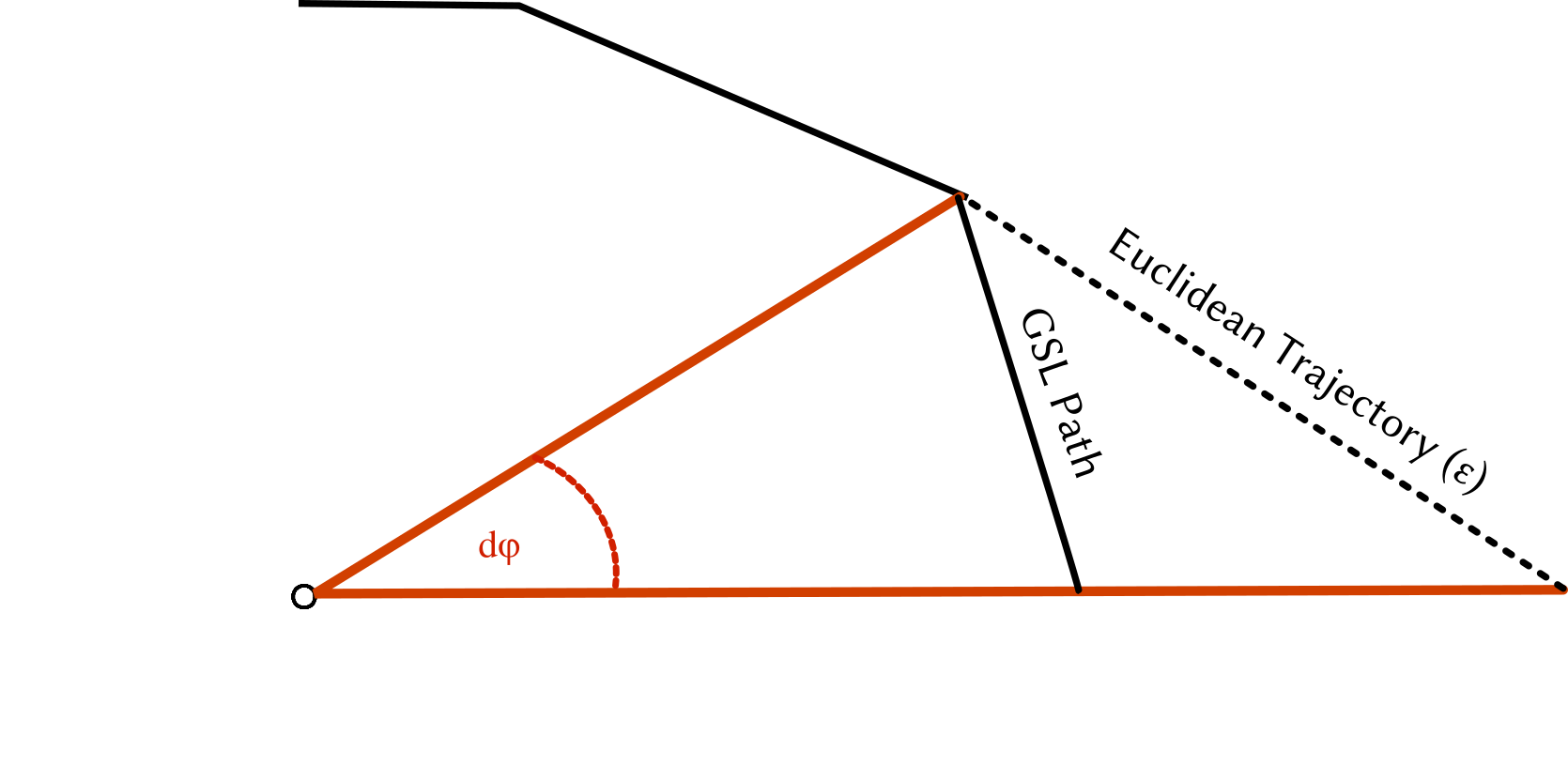}
  \caption{We use Euclidean geometry to compute $d\phi$, then rely on GSL to stencil the ray's path until the simulation progresses by $d\phi$. The point at the center is the origin of a black hole.}
  \label{fig:nextpoint}
\end{figure}

As for how to compute these segments, the GNU Scientific Library's (GSL) suite \cite{10.5555/1538674} of ODE approximation tools found significant utility. GSL allows us to define our equations as C functions, and our particular ray by a set of initial conditions. From there, we can use GSL to compute a series of discrete points on the ray governed by a timestep of our choosing. For initial conditions, we need to describe, in polar coordinates, an initial point on the ray and a single successor point, which was computed using the linear direction, as we are casting rays far from the black hole's influence. As for the \textit{timestep} (as it is commonly known in the GSL documentation), we allowed program users to control the distance between computed points, which we will refer to as $\varepsilon$, i.e. the \textit{discretization} or \textit{step} of the simulation. Lower values of $\varepsilon$ result in more accurate trajectories, but also require more computed points. Fig~\ref{fig:nextpoint} depicts an example of how these paths are computed.

\subsection{Domain Decomposition}

The essential element of speeding up a problem with Message Passing Interface (MPI) \cite{mpi41} is often to attempt to decompose the problem space into subproblems which can be computed on an individual node. In our case, our problem space is the image we are attempting to render. The most simple solution is to decompose along the scanlines, and render distinct scanlines on distinct nodes. In effect, one would attempt to assign a "band" to each process to render against.

\subsection{Use of OpenMP}
In the context of a single band, it is quite easy to use OpenMP \cite{660313_OMP} to accelerate rendering, as the problem can be decomposed along the individual scanlines. This was quite helpful, as MPI was best used to distribute the problem across nodes, and OpenMP was useful for distributing the problem across cores within the node.

\subsection{Code Availability}
Considering all the elements described in the previous sections, we developed our
source code in C++ which is available under a GPL-v2 license, in the following
GitHub repository:
	\url{https://github.com/liamnaddell/BHRaytracer}.


\section{Results}
\label{sec:results}

\subsection{A selection of renders}

\begin{figure*}[h]
  \centering
 \begin{minipage}{0.48\linewidth}
  \includegraphics[width=\linewidth]{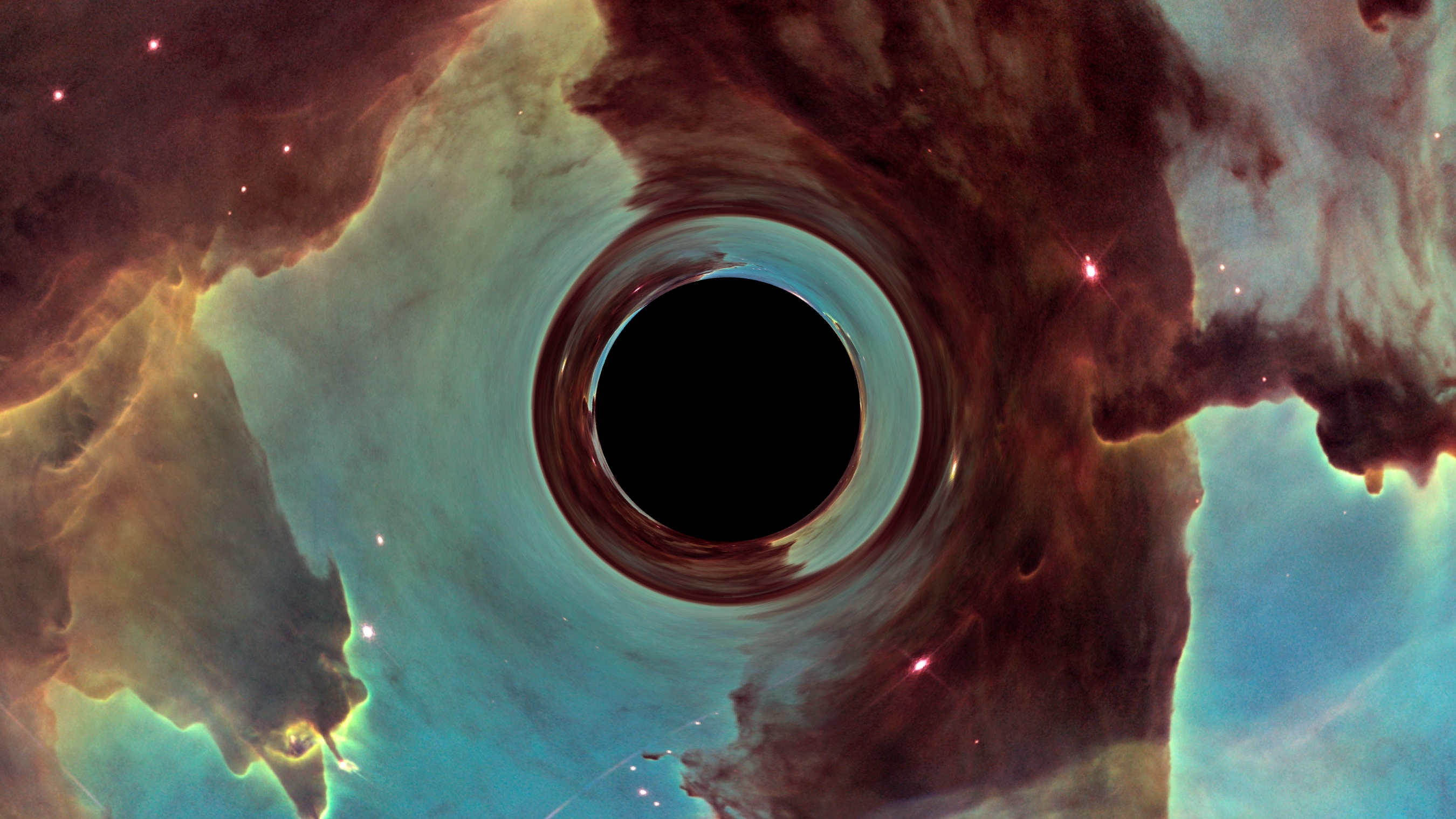}
  \caption{Image generated from a portion of the Eagle Nebula M16 downloaded from \cite{esa-pillars}. }
  \label{fig:eagle}
 \end{minipage}
  \hspace{.01\linewidth}
 \begin{minipage}{0.48\linewidth}
  \centering
  \includegraphics[width=\linewidth]{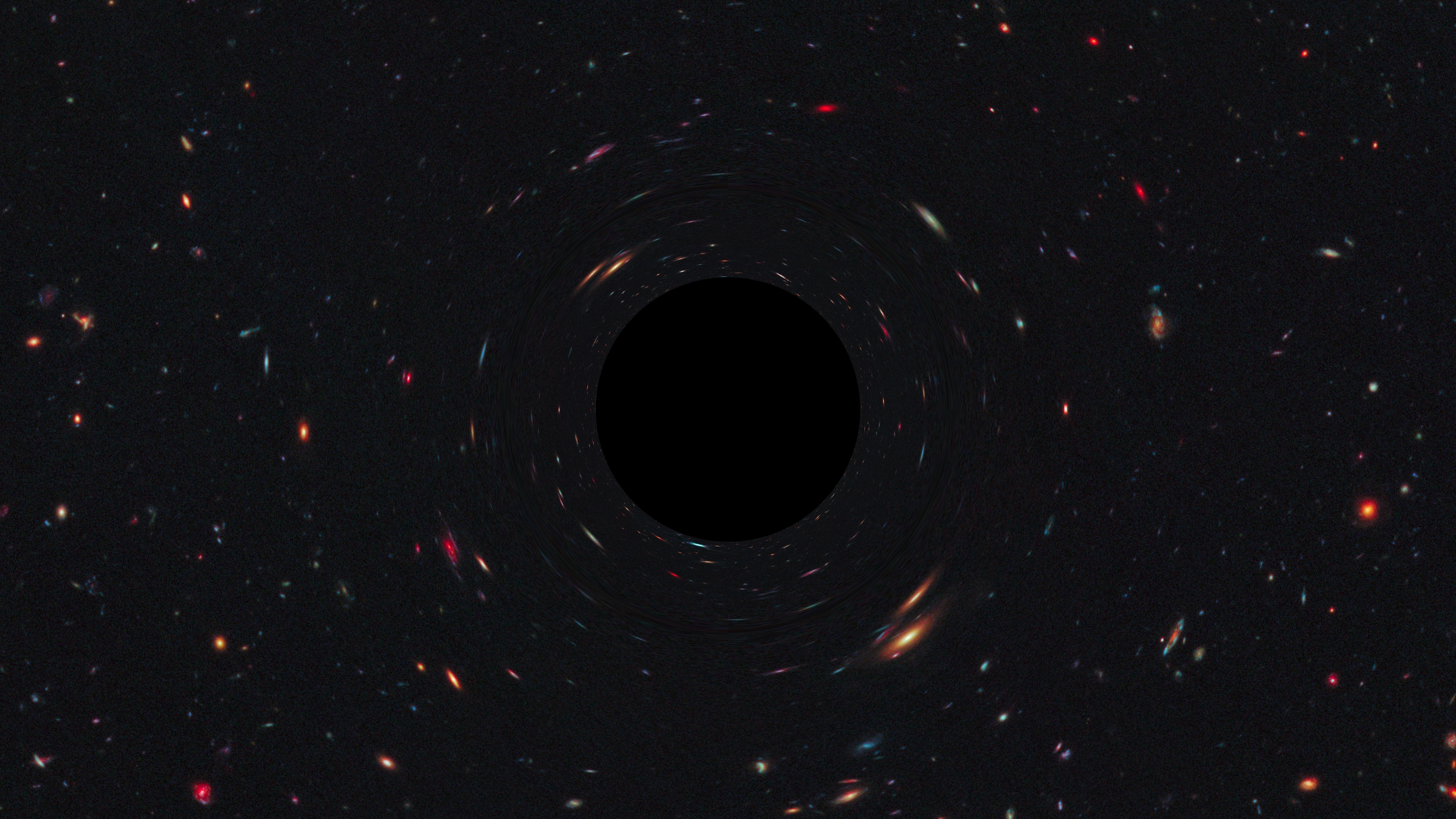}
  \caption{A second sample image from our ray tracer,
	in this case from the CANDELS project\cite{candels-nasa_svs}.
    }
  \label{fig:starry}
 \end{minipage}
\end{figure*}

This section includes results of images rendered employing our ray tracer's implementation.
The background images for these renders were taken from NASA's Scientific Visualization Studio and the ESA/Nasa Hubble Mission.
We present two images -- Figs.~\ref{fig:eagle} and \ref{fig:starry}, it is easy to see the effects of gravitational lensing, as well as the clear Einstein ring. 

\subsection{Scaling analysis}


For this problem, it is possible to consider different \textit{metrics} to analyze the scaling behavior of a given implementation.
For instance, one could consider increasing the size of the background image to stress image parsing;
or instead increase the number of rendered pixels;
or decrease $\varepsilon$ yielding to more calls into GSL;
or increase samples-per-pixel which controls antialiasing.

When increasing the size of the background image, we estimate that this increases the time spent outside the parallel region, as the background image has to be parsed by each MPI process.
Because of this, we consider that this does not present significant interest for scaling analysis.
In contrast, the number of rendered pixels, and the number of cores used present useful scaling parameters, since they correspond most directly to the amount of work needing to be done, and the number of workers to do it.

\begin{figure*}[h]
  \centering
 \begin{minipage}{0.45\linewidth}
  \includegraphics[width=\linewidth]{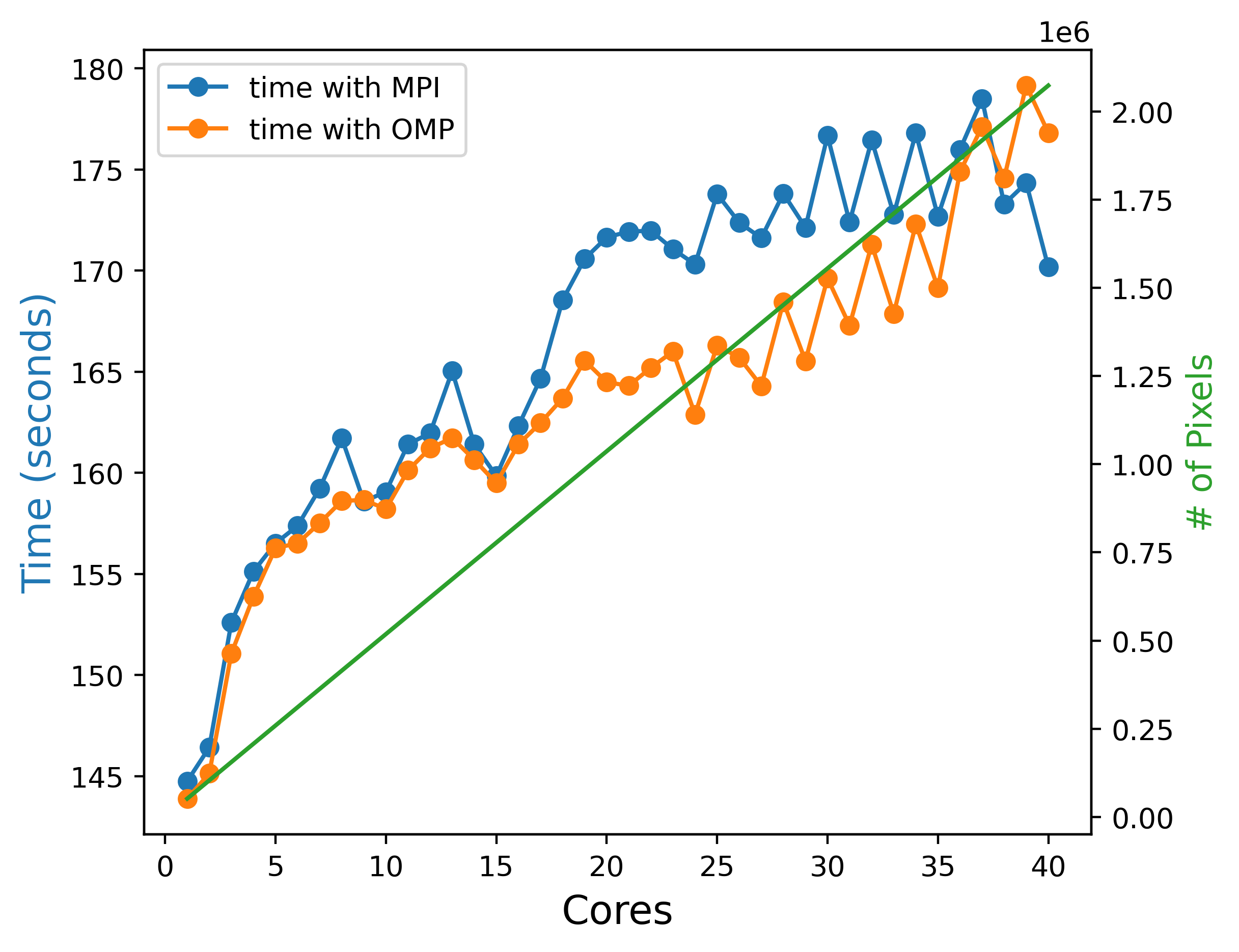}
 \caption{Weak scaling analysis where MPI process/OMP thread count and image width are increased in tandem.}
	  \Description{As core count and image width increase linearly, time increases linearly }
    \label{fig:mpi_weak2}
    \end{minipage}
  \hspace{.05\linewidth}
  \begin{minipage}{0.45\linewidth}
      \includegraphics[width=\linewidth]{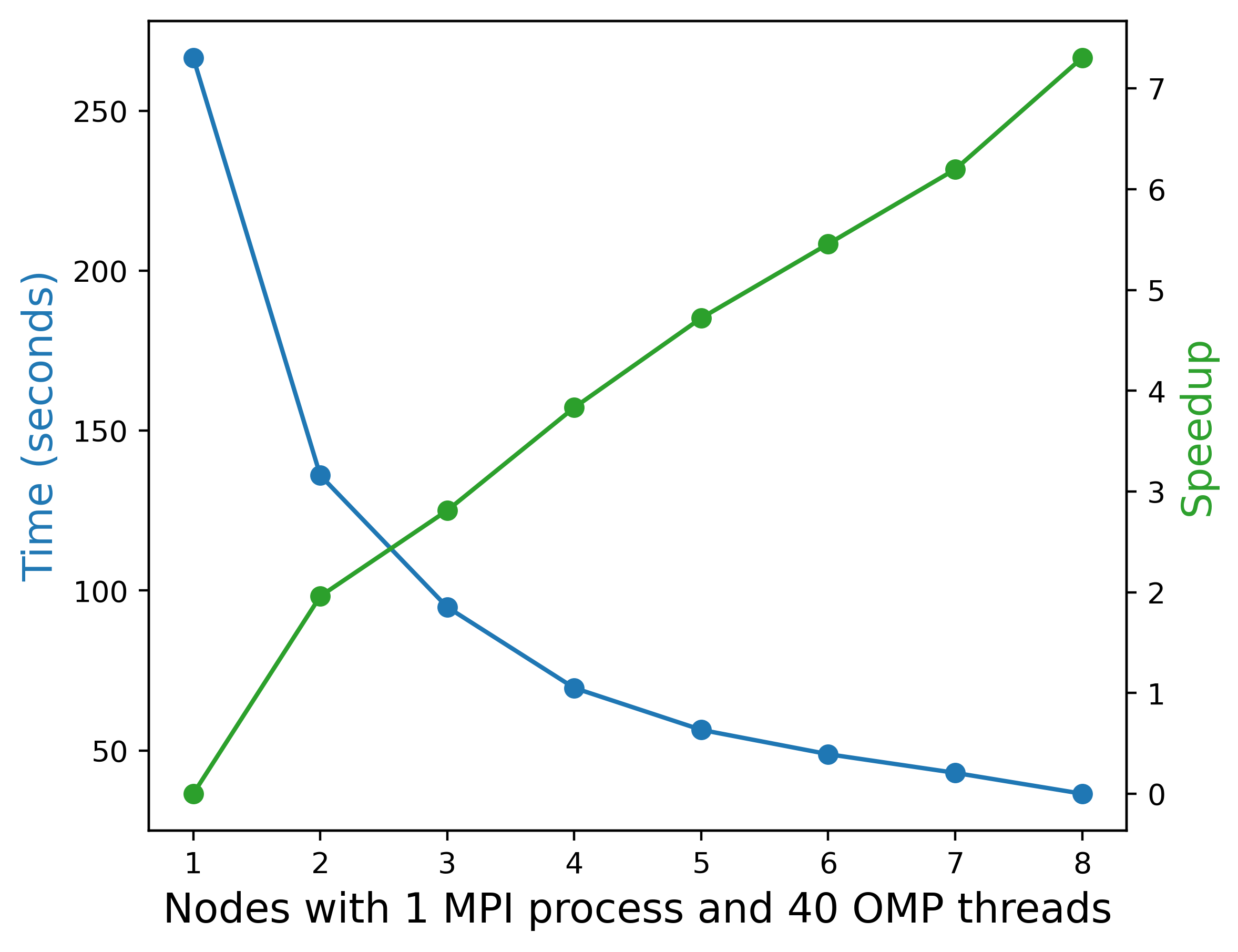}
      \caption{Multi-node strong scaling analysis shows good speedup across 8 nodes.}
	  \Description{Time still converges to serial fraction even on a mutlinode system }
        \label{mpi_strong_multinode}
    \end{minipage}
  \begin{minipage}{0.45\linewidth}
      \includegraphics[width=\linewidth]{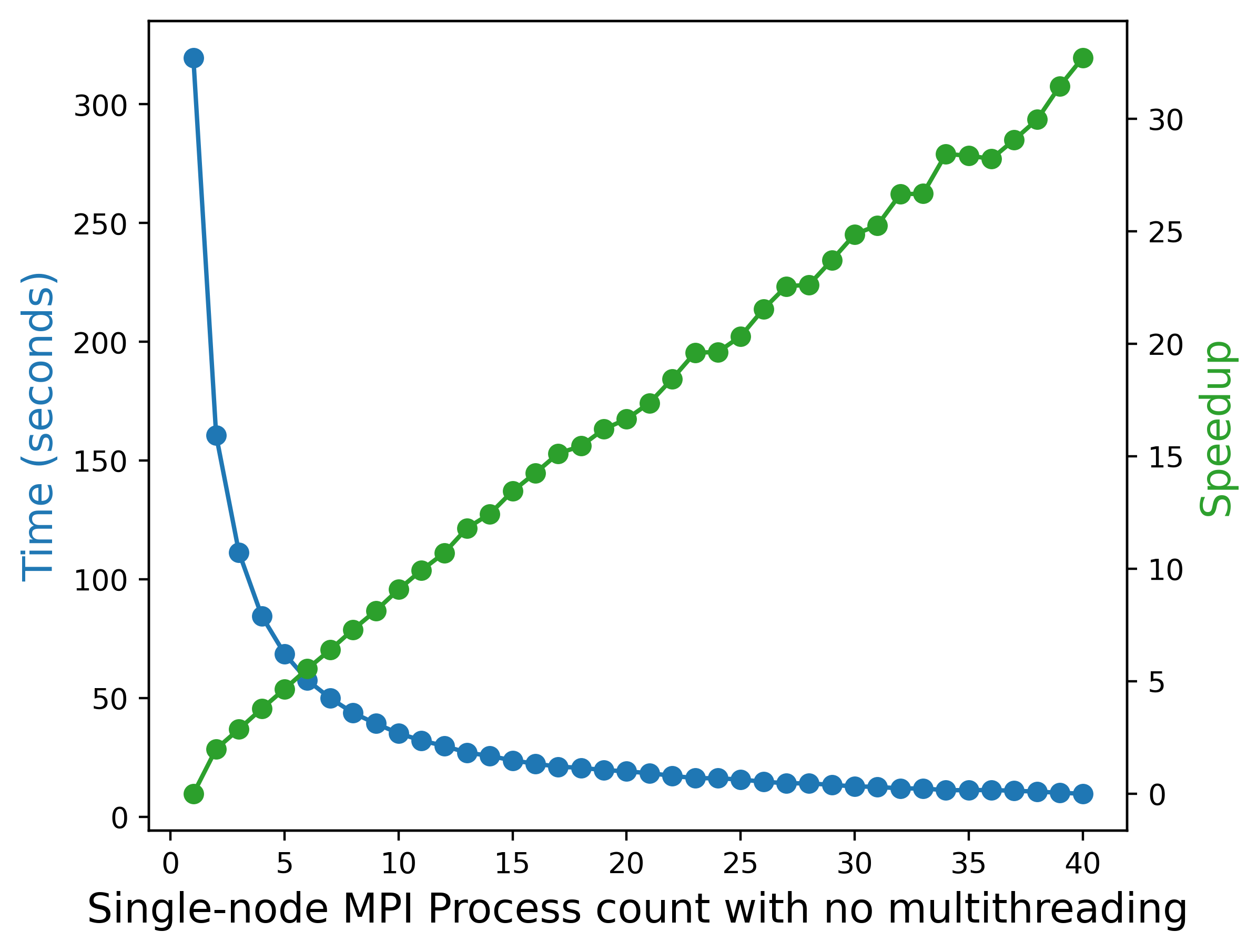}
	  \caption{Strong scaling analysis as MPI process count is increased. }
      \Description{Speedup is nearly linear and time plateaus}
      \label{fig:mpi_strong}
    \end{minipage}
  \hspace{.05\linewidth}
 \begin{minipage}{0.45\linewidth}
  \includegraphics[width=\linewidth]{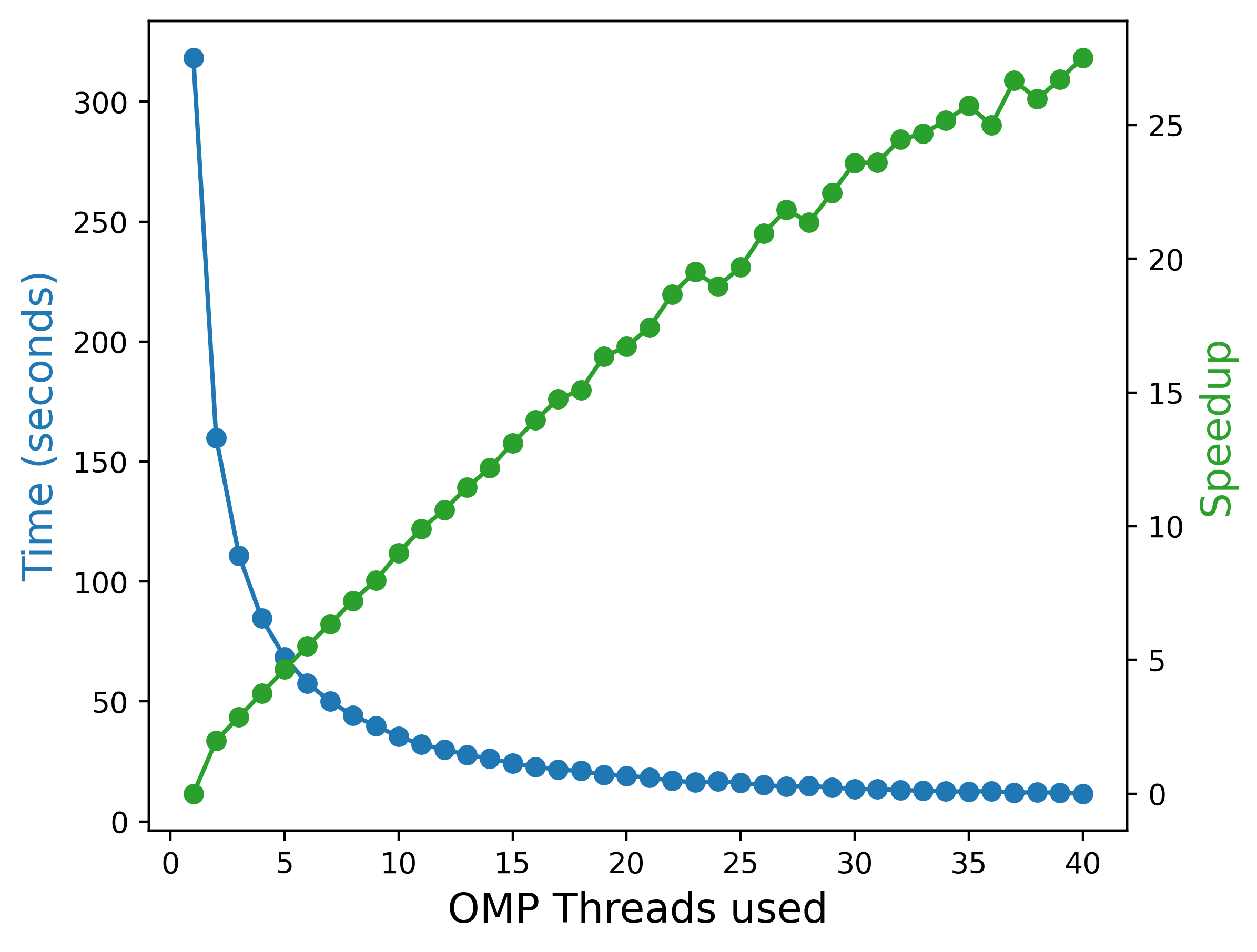}
  \caption{Strong scaling analysis with OpenMP threads -- MPI and OpenMP have similar performance.}
    \Description{Time converges to the serial fraction, and speedup decreases with increased core count}
    \label{fig:omp_strong}
    \end{minipage}
\end{figure*}

Fig.~\ref{fig:mpi_weak2} shows a weak scaling analysis where we linearly increase the number of pixels rendered and the single-node cores utilized in tandem. We observe that both OpenMP and MPI start at around 145 seconds, and increase logarithmically before plateauing roughly around 170-185 seconds per run. While ideally, we would observe a consistent time which does not increase, these results show relatively stable performance. We also witness OpenMP outperforming MPI most of the time, which is expected, since MPI requires transferring pixel data between processes.

We observe something similar with the strong scaling results in Figs.~\ref{fig:mpi_strong} and ~\ref{fig:omp_strong}. These results were run by increasing the number of cores used with a fixed workload. These results show an almost perfect convergence down to the serial fraction with similarly competitive results between OpenMP and MPI.
It is worth noticing as well when comparing Figs.~\ref{fig:mpi_strong} and \ref{fig:omp_strong}
 that although the runtimes are within a few seconds, the difference in time can be appreciated by comparing relative speedups.

All the results and analysis presented in this section were obtained
running our code on the Niagara supercomputer \cite{10.1145/3332186.3332195},
which is a large homogeneous cluster composed by 2,024 Lenovo SD530 servers
each with 40 Intel "Skylake"/"CascadeLake" cores working at approx. 2.4/2.5 GHz.
Each result is an average of 5 runs with identical parameters, and each MPI process is allocated 1 rank.


\section{Discussion}
\label{sec:disc}

\subsection{Educational Value}

This ray tracer implementation was originally developed as a final project for
an upper year specialized topics course in Computer Science, where the main
elements in the course were the discussion of High-Performance Computing
techniques.
As a rich computational project, it combines elements from multiple disciplines, specifically
physics simulations (both through the use of a ray tracer, and through the simulation of gravitational distortion),
solutions to differential equations via approximations with GSL,
path stenciling, image parsing,
parallel and distributed computing with MPI and OpenMP, and domain decomposition.
These factors make the project compelling as an educational tool.


\section{Conclusions}
\label{sec:concl}

Gravitational lensing is a beautiful phenomenon, both when witnessed from space, and when digitally rendered.
This paper explored a relatively simple method for creating a ray traced image which shows gravitational lensing, and how to accelerate the process using shared and distributed memory computing.
One may argue that the ray-tracing problem has been solved,
however, taking a start-from-scratch approach, can be useful to highlight important and fundamental aspects of the basic techniques and implementations used.
Furthermore, an approach such as this offers multiple profound opportunities in the areas of teaching and education, both with regards to general scientific and High-Performance computing techniques.


\section*{Acknowledgment}

Computations were performed on the Niagara supercomputer and the Teach cluster at the SciNet HPC Consortium. SciNet is funded by Innovation, Science and Economic Development Canada; the Digital Research Alliance of Canada; the Ontario Research Fund: Research Excellence; and the University of Toronto.
We are thankful to SciNet for allowing us to use their HPC systems.
In particular, to Danny Gruner, Ramses van Zon, Vladimir Slavnic and
Norbert Krawiec, for their continuous support and hospitality.

\bibliographystyle{ACM-Reference-Format}
\bibliography{refs}

\end{document}